\documentclass[12pt]{article}
\usepackage[dvips]{graphicx}
\usepackage{amssymb}
\title{Regularized Renormalization Group Reduction of Symplectic Maps}
\author{Shin-itiro Goto\footnote{e-mail:~sgoto@allegro.phys.nagoya-u.ac.jp}
\quad and \quad Kazuhiro Nozaki\\
{\it Department of Physics,Nagoya University,Nagoya 464-8602,Japan}}
\date{}
\makeindex

\begin{document}
\maketitle

\newcommand{\beq}{\begin{equation}}
\newcommand{\beqa}{\begin{eqnarray}}
\newcommand{\eeq}{\end{equation}}
\newcommand{\eeqa}{\end{eqnarray}}
\newcommand{\non}{\nonumber}
\newcommand{\lb}{\label}
\newcommand{\fr}[1]{(\ref{#1})}
\newcommand{\cc}{\mbox{c.c.}}
\newcommand{\nr}{\mbox{n.r.}}
\newcommand{\tx}{\widetilde{x}}
\newcommand{\tg}{\widetilde{g}}
\newcommand{\hx}{\widehat{x}}
\newcommand{\tA}{\widetilde A}
\newcommand{\tB}{\widetilde B}
\newcommand{\tK}{\widetilde K}
\newcommand{\tc}{\widetilde c}
\newcommand{\tAc}{{\widetilde A}^{*}}
\newcommand{\tphi}{{\widetilde \phi}}
\newcommand{\btA}{\mbox{\boldmath {$\widetilde A$}}}
\newcommand{\bA}{\mbox{\boldmath {$A$}}}
\newcommand{\bC}{\mbox{\boldmath {$C$}}}
\newcommand{\bu}{\mbox{\boldmath {$u$}}}
\newcommand{\bN}{\mbox{\boldmath {$N$}}}
\newcommand{\bZ}{\mbox{\boldmath {$Z$}}}
\newcommand{\bR}{\mbox{\boldmath {$R$}}}
\begin{abstract}
By means of the perturbative renormalization group method, we
study a long-time behaviour of some symplectic discrete maps near
elliptic and hyperbolic fixed points.  It is shown that a naive
renormalization group (RG) map breaks
the symplectic symmetry and fails to describe a long-time behaviour.
In order to preserve the symplectic
symmetry, we present a regularization procedure, which gives
a regularized symplectic RG map describing an approximate
long-time behaviour succesfully.
\end{abstract}
\section{Introduction}
\qquad
There has been a long history to study an asymptotic solution of
Hamiltonian flows by means of singular perturbation methods such as
the averaging method and the method of multiple time-scales.
A Hamiltonian flow can be reduced to a symplectic discrete map
called the Poincare map,
which has the lower dimension than the original flow and is, therefore,
extensively studied. However, neither the averaging method
nor the method of  multiple time-scales may  be immediately applicable to
symplectic maps.
The perturbative renormalization group (RG) method developed recently
may be a useful tool to tackle asymptotic
behaviours of discrete maps as well as flows.
The original RG method is an asymptotic singular perturbation technique
developed for diffential equations \cite{CGO96}. Secular or divergent terms
of  perturbation solutions of  differential equations are
removed by renormalizing integral constants of the lowest order solution.
The RG method is reformulated on the basis of a naive renormalization
transformation and the Lie group \cite{GMN99}.
This reformulated RG method based on the Lie group is easy to apply to
discrete systems, by which asymptotic expansions of unstable
manifolds of some chaotic discrete systems are obtained \cite{GN99}.
The extension of the RG method to discrete symplectic systems is
not trivial because the symplectic struture is easily broken
in naive renormalization group equations (maps)
as shown in this paper, while the application of the RG method to
Hamiltonian flows does not cause such a problem as the broken
symplectic symmetry except a special case \cite{YN98}.
The application of the RG method to some non-symplectic discrete
systems was tried in the framework of the envelope method \cite{KM98}.
Howevever, the method,if applied to a symplectic map, would give only
a naive RG map which breaks the symplectic symmetry.\\

The main purpose of the present paper is to present
an RG procedure to preserve the symplectic structure in
RG maps and to obtain symplectic RG maps. In this paper, this procedure
 is called a regularized RG procedure, which consists of the following
 two steps.
First,using the reformulated RG method \cite{GMN99}, we get a naive
RG map near elliptic and hyperbolic fixed points
of some symplectic discrete systems.  The naive map preserves the
symplectic symmetry only approximately and fails to describe
a long-time behaviour of the original maps.
Second, in order to recover the symplectic symmetry,
we introduce a process of
``exponentiation'' which yields a symplectic RG map.
This process and a symplectic RG map thus obtained are
called,respectively, regularization of an RG map and a regularized
RG map.\\
In section 2, a long-time behaviour
of a simple linear map is analyzed  in order to elucidate the
broken symplectic symmetry in a naive RG map and our regularization
process.
In section 3 , regularized (symplectic) RG maps are obtained near
elliptic and hyperbolic fixed points of a two-dimensional nonlinear
symplectic map. In section 4, the regularized RG procedure is
applied to a
four-dimensional symplectic map, of which  elliptic-elliptic fixed point
has an irrational frequency ratio.

\section{Linear symplectic map}
\qquad 
It may be instructive to  analyze a linear symplectic map, which
is exactly solvable,
$\bR^2\ni (x_n,y_n)\mapsto (x_{n+1},y_{n+1})$ :
\beqa
x_{n+1}&=&x_n+y_{n+1},\non\\
y_{n+1}&=&y_n-Jx_n+2\epsilon Jx_n,\lb{lin-1}
\eeqa
where $n \in\bZ $ and
$\epsilon$ is a small parameter.
The map \fr{lin-1}  has a elliptic fixed point at the
origin (0,0) for $0<J<2$.
Eliminating $y$ from \fr{lin-1}
 ,we obtain a second order difference equation:
\beq
Lx_n\equiv x_{n+1}-2\cos(\theta) x_n+x_{n-1}=\epsilon 2 J x_n,\lb{lin-ell}
\eeq
where $\cos\theta=1-J/2$ and $0<J<2$ is assumed.
The linear map \fr{lin-ell} has the following exact solution $x_n^E$ :
\beqa
x_n^E&=&A\exp\bigg(i\arccos(\cos\theta+\epsilon J)n\bigg)+\cc,\non\\
     &=&A\exp\Bigg[i\Bigg(\theta +\epsilon\frac{-J }{\sin\theta}
                +\epsilon^2\frac{-\cos\theta}{2\sin\theta}
                          \bigg(\frac{J}{\sin\theta}\bigg)^2
                          +\cdots \Bigg)n\Bigg]+\cc,\lb{lin-ell_exact}
\eeqa
where $A(\in\bC)$ is a complex ``integral'' constant
and $\cc$ stands for the complex conjugate to the preceding terms.\\
Let us derive an asymptotic solution of the map \fr{lin-1}
for small $\epsilon$ by means of the RG method.
Substituting the expansion :
\beq
x_n=x_n^{(0)}+\epsilon x_n^{(1)}+\epsilon^2x_n^{(2)}+\cdots, \non
\eeq
 into Eq.\fr{lin-ell}, we have
$$
Lx_n^{(0)}=0,\quad Lx_n^{(1)}=2Jx_n^{(0)},\quad Lx_n^{(2)}=2Jx_n^{(1)}
\cdots.
$$
and
\beqa
x_n^{(0)}&=&A\exp(i\theta n)+\cc,\non\\
x_n^{(1)}&=&\frac{JA}{i\sin\theta}n\exp(i\theta n)+\cc,\non\\
x_n^{(2)}&=&\frac{-J^2A}{2\sin^2\theta}\bigg(
     n^2+i\frac{\cos\theta}{\sin\theta}n\bigg)
\exp(i\theta n)+\cc,\non
\eeqa
where $A(\in\bC)$ is a complex constant.
To remove secular terms ($\propto n, n^2$),
we introduce a renormalization transformation
$A\mapsto \tA(n)$ \cite{GMN99}:
\beq
\tA(n)\equiv A+\epsilon\frac{JA}{i\sin\theta}n
             +\epsilon^2\frac{-J^2A}{2\sin^2\theta}\bigg(
                 n^2+i\frac{\cos\theta}{\sin\theta}n\bigg)
             +{\cal O}(\epsilon^3),\lb{lin-ell_tA}
\eeq
A discrete version of the RG equation is just the first order
difference equation of $\tA(n)$, whose local solution is given by
Eq.\fr{lin-ell_tA}.
From Eq.\fr{lin-ell_tA}, we have
\beq
\tA(n+1)-\tA(n)=\Bigg(-i\epsilon\frac{J}{\sin\theta}
                     -\epsilon^2\frac{J^2}{2\sin^2\theta}
                 \bigg(2n+1+i\frac{\cos\theta}{\sin\theta}\bigg)
                 \Bigg)A+{\cal O}(\epsilon^3),\lb{lin-nrg1}
\eeq
where  $A$  should be
expressed in terms of $\tA(n)$. This is done by taking
the inversion of the renormalization transformation \fr{lin-ell_tA}
iteratively.
\beq
A=\bigg(1+i\epsilon\frac{Jn}{\sin\theta}+{\cal O}(\epsilon^2)\bigg)\tA(n).
\lb{invA}
\eeq
Substituting \fr{invA} into \fr{lin-nrg1}, we obtain
the following RG equation (RG map) up to ${\cal O}(\epsilon^2)$.
\beq
\tA(n+1)=\Bigg(1+\frac{-i\epsilon J}{\sin\theta}
          +\frac{1}{2!}\bigg(\frac{-i\epsilon J}{\sin\theta}\bigg)^2
          -i\epsilon^2\frac{J^2\cos\theta}{2\sin^3\theta}\Bigg)\tA(n)
+{\cal O}(\epsilon^3),\lb{lin-ell_nrg}
\eeq
of which solution is
\beq
\tA(n)=\Bigg(1+\frac{-i\epsilon J}{\sin\theta}
          +\frac{1}{2!}\bigg(\frac{-i\epsilon J}{\sin\theta}\bigg)^2
          -i\epsilon^2\frac{J^2\cos\theta}{2\sin^3\theta}
          +{\cal O}(\epsilon^3)\Bigg)^n\tA(0)
.\lb{lin-sol}
\eeq
On the other hand, from Eq. \fr{lin-ell_exact},
we have $\tA(n)$ exactly as
\beq
\tA(n)=\tA(0)\exp\Bigg[i\Bigg(\epsilon\frac{-J }{\sin\theta}
                -\epsilon^2\frac{\cos\theta}{2\sin\theta}
                          \bigg(\frac{J}{\sin\theta}\bigg)^2
                          +\cdots \Bigg)n\Bigg],\lb{lin-extsol}
\eeq
Notice that $|\tA|^2$ is an exact constant of motion  while
it is merely an approximate conserved quantity of
the (trancated) RG map \fr{lin-ell_nrg}.
The symplectic structure is also not exactly preserved  in the
RG map, that is, for the truncated RG map \fr{lin-ell_nrg}
up to ${\cal O}(\epsilon^k)$, we have
$$
d\tA(n+1)\land d\tAc(n+1)-d\tA(n)\land d\tAc(n)={\cal O}(\epsilon^{(k+1)})
\ne 0,
$$
where $k=1,2,\cdots$. $\tA^{*}$ is complex conjugate to $\tA$ and
should also be a canonical conjugate to $\tA$ \cite{nt-flow}.
Although this fault of the RG map vanishes in the limit of small $\epsilon$,
it is intorerable as shown in Fig.1 where
we depict a long time behaviour of the solution for small but
finite $\epsilon$.
In fact, the truncated RG map becomes a system with
small but finite dissipation .
In order to remedy a fault like this, we take advantage of a crucial
observation that the coefficient of $\tA(n)$  in
\fr{lin-ell_nrg} can be modified as
\beqa
& &1+\frac{-i\epsilon J}{\sin\theta}
          +\frac{1}{2!}\bigg(\frac{-i\epsilon J}{\sin\theta}\bigg)^2
          -i\epsilon^2\frac{J^2\cos\theta}{2\sin^3\theta} \non \\
&=&\exp\Bigg(\frac{-i\epsilon J}{\sin\theta}
          -i\epsilon^2\frac{J^2\cos\theta}{2\sin^3\theta}\Bigg)
  +{\cal O}(\epsilon^3). \non
\eeqa
Using this modified coefficient, the symplectic symmetry in the
truncated RG map is recovered as
\beq
\tA(n+1)=\exp\Bigg(\frac{-i\epsilon J}{\sin\theta}
          -i\epsilon^2\frac{J^2\cos\theta}{2\sin^3\theta}\Bigg)\tA(n)
,\lb{lin-ell_erg}
\eeq
which also has an exact conserved quntity $|\tA|$.
This process is nothing but ``exponentiation'' of  the coefficient
of $\tA(n)$ and is called  regularization of an RG map.
It is easy to see that the solution of the regularized RG map
\fr{lin-ell_erg} coincides with the asymptotic expansion of
the exact solution \fr{lin-extsol} and describes a long time behaviour
up to $ n\sim {\cal O}(\epsilon^{-2}) $ in the present approximation.
In Fig.(1), trajectories constructed from the naive RG map \fr{lin-ell_nrg}
and the regularized RG map \fr{lin-ell_erg} are depicted to be compared
to an ``exact'' trajectry of the original map \fr{lin-1} 

\begin{figure}[]
\begin{center}
\includegraphics[width=12cm]{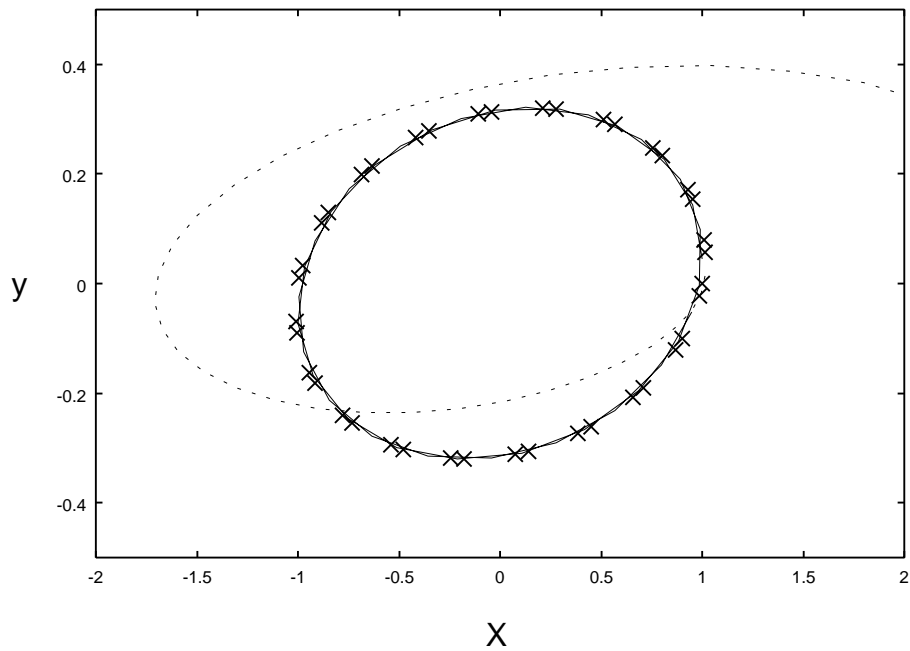}
\caption{Trajectries constructed from the naive RG map (broken line),
the regularized RG map(solid line) and an ``exact'' trajectry of the
original map \fr{lin-1} (crosses) for $J=0.2,\epsilon=0.25$ and an
initial condition $(x_0=1.0,y_0=0.0)$.}
\end{center}
\end{figure}

\pagebreak

\section{Two-dimensional Non-linear Symplectic Map}
\subsection{Elliptic Fixed Point}
\qquad
Let us analyze a weakly non-linear symplectic map\\
$\bR^2\ni (x_n,y_n)\mapsto (x_{n+1},y_{n+1})$ :
\beqa
x_{n+1}&=&x_n+y_{n+1},\non\\
y_{n+1}&=&y_n-Jx_n+2\epsilon Jx_n^3,\lb{nonmap}
\eeqa
or
\beq
Lx_n=\epsilon 2 J x_n^3,\lb{ell}
\eeq
where $\epsilon $ is a small parameter, $L$ is defined in \fr{lin-ell}
and  $0<J<2$ is assumed so that Eq.\fr{ell} has a elliptic fixed point at
the origin (0,0).
Expanding $x_n$ as a power series of $\epsilon$
\beq
x_n=x_n^{(0)}+\epsilon x_n^{(1)}+\epsilon^2x_n^{(2)}+\cdots,\lb{expan}
\eeq
we have
$$
Lx_n^{(0)}=0,\quad Lx_n^{(1)}=2Jx_n^3,\quad Lx_n^{(2)}=6Jx_n^{(0)}x_n^{(1)},
\cdots ,
$$
and solutions of the perturbed equations to ${\cal O}(\epsilon^2)$ are
given by
\beqa
x_n^{(0)}&=&A\exp (i\theta n)+\cc ,\non\\
x_n^{(1)}&=&\alpha_1 A^3\exp(3i \theta n)+i\alpha_{1R}|A|^2An\exp(i\theta n)
+\cc ,\non\\
x_n^{(2)}&=&i\alpha_1\alpha_{1R}|A|^4An\exp (i\theta n)
-\frac{\alpha_{1R}^2}{2}|A|^4A\Bigg(n^2+i\frac{\cos\theta}{\sin\theta}n\Bigg
)\exp(i\theta n)\non\\
&&+3i\alpha_1\alpha_{1R}|A|^2 A^3
\Bigg(n-i\frac{\sin(3\theta)}{\cos3\theta-\cos\theta}\Bigg)
\exp(3i\theta n)\non\\
&&+6\alpha_1^2|A|^2A^3\exp(3i\theta n)
+\frac{3J\alpha_1A^5}{\cos 5\theta -\cos\theta}\exp(5i\theta)+\cc,\non
\eeqa
where
$$
\alpha_1\equiv\frac{J}{\cos 3\theta-\cos\theta},\quad
\alpha_{1R}\equiv\frac{-3J}{\sin\theta},
$$
 and $A(\in \bC)$ is a complex integral constant.
In order to remove secular terms in the coefficient of
the fundamental harmonic ($\exp(i\theta n)$),
we introduce a renormalization transformation
$A\mapsto \tA(n)$:
\beqa
\tA(n)&=&A+\epsilon (i\alpha_{1R}|A|^2An)\non\\
      &&  +\epsilon^2 \Bigg(\frac{-\alpha_{1R}^2}{2}|A|^4A
      \bigg(n^2+i\frac{\cos\theta}{\sin\theta}n\bigg)
+i\alpha_1\alpha_{1R}|A|^4An\Bigg). \lb{ell_tA}
\eeqa
Following the same procedure as that in the preceding section, we derive
a naive RG map from Eq.\fr{ell_tA}:
\beqa
\tA(n+1)-\tA(n)&=&\epsilon i \alpha_{1R}|\tA(n)|^2\tA(n)\non\\
&&+\epsilon^2 \Bigg(
  \frac{-\alpha_{1R}^2}{2}|\tA(n)|^4\tA(n)
     \bigg(1+i\frac{\cos\theta}{\sin\theta}\bigg)\non\\
&&  +i\alpha_1\alpha_{1R}|\tA(n)|^4\tA(n)
              \Bigg),\lb{ell_nrg}
\eeqa
which breaks the symplectic symmetry and
does not have a constant of motion.
To recover the symplectic symmetry,
we regularize the naive RG map \fr{ell_nrg} by
``exponentiating'' the coefficient of $\tA(n)$, that is,
\beqa
\tA(n+1)&=&\tA(n)\exp\Bigg(
   i\epsilon|\tA(n)|^2\alpha_{1R}\non\\
   &&+i\epsilon^2|\tA(n)|^4
   \bigg(\frac{-\cos\theta}{2\sin\theta}\alpha_{1R}^2+
      \alpha_{1}\alpha_{1R}\bigg)
\Bigg),\lb{ell_erg}
\eeqa
which of course agrees with Eq.\fr{ell_nrg} to ${\cal O}(\epsilon^2)$.
It is easy to see that Eq.\fr{ell_erg} has the symplectic symmetry
$$d\tA(n+1)\land d\tAc(n+1)=d\tA(n)\land d\tAc(n),$$
and a constant of motion $|\tA|$.
Introducing the polar expression $\tA(n)=|\tA(n)|\exp(i\tphi(n))$,
the regularized RG map is reduced to
a simple phase equation given by
\beqa
\tphi (n+1)&=&\tphi (n)+ \Bigg(
   \epsilon|\tA(0)|^2\alpha_{1R}\non\\
&&+\epsilon^2|\tA(0)|^4
   \bigg(\frac{-\cos\theta}{2\sin\theta}\alpha_{1R}^2+
      \alpha_{1}\alpha_{1R}\bigg)
\Bigg),
\eeqa
Thus, the regularized RG map \fr{ell_erg} is exactly solvable in contrast
to the naive RG map \fr{ell_nrg} which do not have a conserved quantity.
It is noted that  secular coefficients of the third harmonic
($\exp(3i\theta n)$) are also summed up to give a renormalized coefficient
$\tA_3(n)$ as:
\beqa
\epsilon\tA_3(n)&\equiv&\epsilon\alpha_1A^3\non\\
             &&+\epsilon^2\Bigg(
    3i\alpha_1\alpha_{1R}|A|^2A^3\bigg(
             n-\frac{i\sin3\theta}{\cos 3\theta-\cos \theta}
                                 \bigg)
 +6\alpha_1^2|A|^2A^3 \Bigg).\lb{ell_tA3}
\eeqa
Substituting an iterative expression of $A$ in terms of $\tA(n)$ obtained
from Eq.\fr{ell_tA} into Eq.\fr{ell_tA3}, we can eliminate secular terms
in the same way as the case of differential equations \cite{GMN99}.\\
As an application of the result, we obtain an approximate but
analytical expression of the rotaion number near the elliptic fixed point.
The definition of the rotation number is
$$
\rho(x_0,y_0)\equiv\lim_{N\to\infty}\frac{1}{2\pi N}\sum_{n=0}^{N}\phi_n,
$$
where $\phi_n$ is an angle between the vectors $(x_n,y_n)$ and
$(x_{n+1},y_{n+1})$ \cite{Aiz89}.
Neglecting effects of the higher hamonics, an approximate expression of
$\rho (x_0,y_0)$ of the present system \fr{nonmap} is gven as following :
\beq
2\pi\rho (x_0,y_0)\approx\theta+\Bigg(
   \epsilon|\tA(0)|^2\alpha_{1R}
+\epsilon^2|\tA(0)|^4
   \bigg(\frac{-\cos\theta}{2\sin\theta}\alpha_{1R}^2+
      \alpha_{1}\alpha_{1R}\bigg)
\Bigg)+\tphi(0),\lb{rotation}
\eeq
where initial values $(x_0,y_0)$ are related to
$|\tA(0)|$ and $\tphi(0)$.
In Fig.(3), the rotaion number $\rho (x_0,y_0)$ given by \fr{rotation}
is depicted for initial values on the half line $0<x_0<0.8 ,\quad y_0=0$
shown in Fig.(2).
The result agrees well with the ``exact'' rotation number obtained
by numerical calculations of the original map \fr{nonmap} until
a new resonance appears near $x_0 \simeq 0.6$.

\begin{figure}
\begin{center}
\includegraphics[width=14cm]{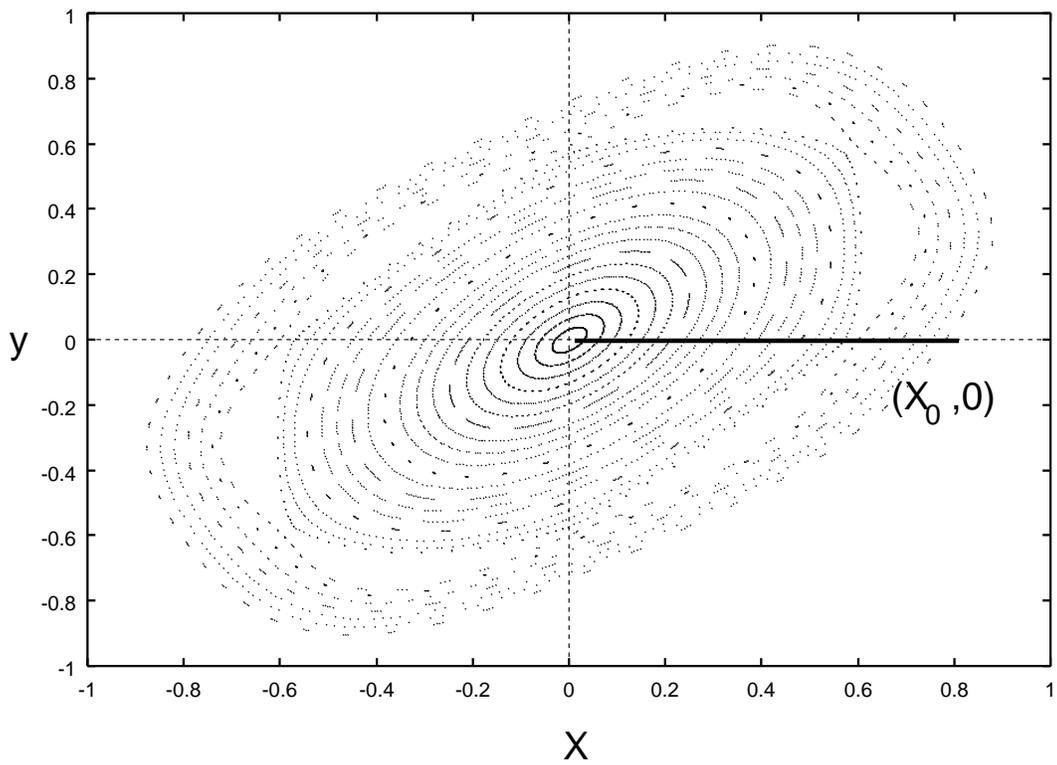}
\caption{A phase portrait of the map \fr{nonmap} for $J=1.2,\quad
\epsilon=0.25$. The solid half line represents initial phase
points $0<x_0<0.8,\quad y_0=0$ of the rotation number $\rho (x_0,y_0)$.}
\end{center}
\end{figure}

\begin{figure}
\begin{center}
\includegraphics[width=14cm]{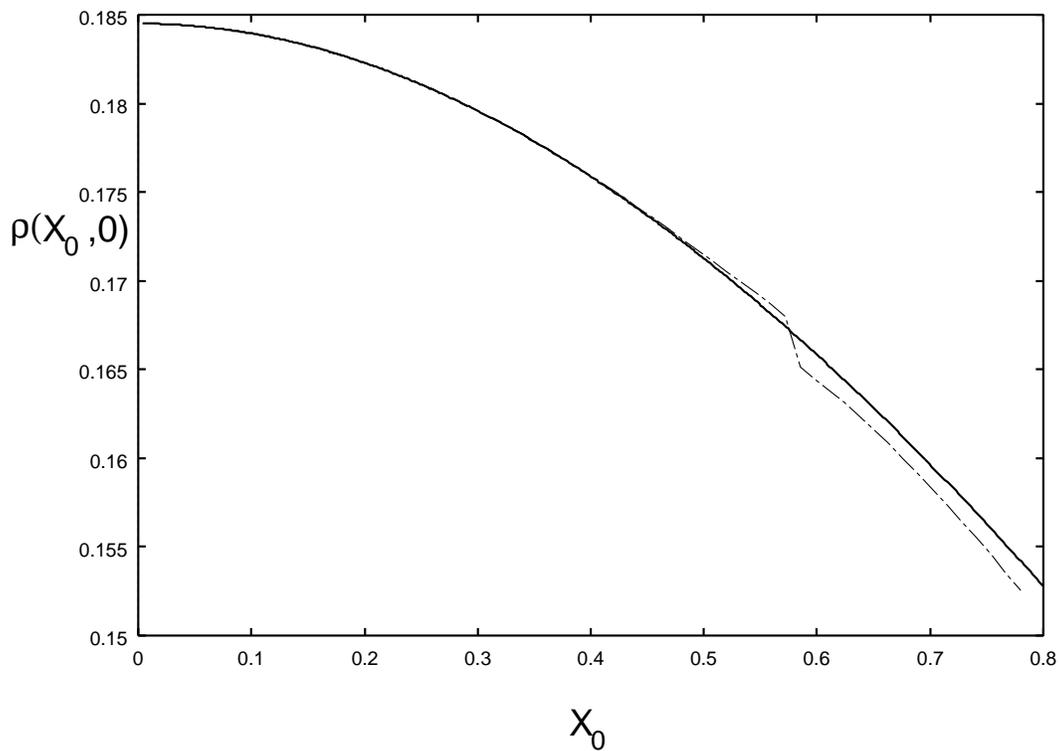}
\caption{The rotation number $\rho (x_0,y_0)$ versus
initial phase points $(x_0,0)$ for $J=1.2,\quad
\epsilon=0.25$. The solid curve represents
the rotation number given by the regularized RG map \fr{rotation},
while the dashed curve is obtained
by numerical calculations of the original map \fr{nonmap}.}
\end{center}
\end{figure}

\pagebreak

\subsection{Hyperbolic Fixed Point}
\qquad 
In the case $J<0$ in \fr{nonmap}, the origin (0,0) is a hyperbolic
fixed point of \fr{nonmap} and Eq.\fr{ell} becomes
\beq
Lx_n=\epsilon 2 J x_n^3,\lb{hyper}
\eeq
where
$$Lx_n\equiv x_{n+1}-2\cosh(\theta) x_n+x_{n-1},\quad
 \cosh\theta=1-J/2 \quad (J<0),$$
and  $K\equiv \exp(\theta)$
 is one of the eigenvalues of the linearized equation at
the origin, that is
$$
K= \frac{1}{2}\bigg(2-J+\sqrt{-4J+J^2}\bigg),\quad \mbox{or}\quad
K= \frac{1}{2}\bigg(2-J-\sqrt{-4J+J^2}\bigg).
$$

Expanding the $x_n$ as \fr{expan},
 we get the following perturbed solution
\beqa
x_n^{(0)}&=&A_{+}K^n+A_{-}K^{-n},\non\\
x_n^{(1)}&=&2J\Bigg(
   \frac{A_{+}^3K^{3n}}{D(K)}
  +\frac{A_{-}^3K^{-3n}}{D(K)}
  +\frac{3A_{+}^2A_{-}}{K-K^{-1}}nK^n
  +\frac{3A_{+}A_{-}^2}{K^{-1}-K}nK^{-n}
    \Bigg),\non\\
x_n^{(2)}&=& 12J^2 \Bigg(
   \frac{3A_{+}^{3}A_{-}^{2}}{2(K-K^{-1})^2}\bigg(n^2-a(K)n\bigg)
  +\frac{A_{+}^{3}A_{-}^{2}}{(K-K^{-1})D(K)}\Bigg) K^n\non\\
&& +12J^2\Bigg(
 \frac{3A_{-}^{3}A_{+}^{2}}{2(K^{-1}-K)^2}\bigg(n^2+a(K)n\bigg)
 +\frac{A_{-}^{3}A_{+}^{2}}{(K^{-1}-K)D(K)}\Bigg) K^{-n}+\nr,\non
\eeqa
where
\beqa
D(K)&\equiv&K^3-(K+K^{-1})+K^{-3}=D(K^{-1})\in \bR ,\non\\
a(K)&\equiv&\frac{K+K^{-1}}{K-K^{-1}}=-a(K^{-1})\in\bR\non.
\eeqa
$A_{+}$ and $A_{-}$ are real ``integral'' constants and
$\nr$ denotes  non-resonant terms which are proportional to
$K^{3n},K^{-3n},K^{5n},K^{-5n}$.
Introducing a renormalization transformation
$A_{+}\mapsto\tA_{+}(n)$ and $A_{-}\mapsto\tA_{-}(n)$
to remove secular terms in the coefficient of $K^{n}$,
we have a set of naive RG equations:
\beqa
\tA_{+}(n+1)&=&\tA_{+}(n)-2\epsilon J
\frac{3\tA_{+}^{2}(n)\tA_{-}(n)}{K-K^{-1}}\non\\
&&+12\epsilon^{2}J^{2}\Bigg(\frac{3\tA_{+}^{3}(n)\tA_{-}^{2}(n)}{2(K-K^{-1})^2}
 \bigg(1-a(K)\bigg)\non\\
&&+\frac{\tA_{+}^{3}(n)\tA_{-}^{2}(n)}{(K-K^{-1})D(K)}\Bigg),
\lb{hyp_rg+}\\
\tA_{-}(n+1)&=&\tA_{-}(n)-2\epsilon
J\frac{3\tA_{-}^{2}(n)\tA_{+}(n)}{K^{-1}-K}\non\\
&&+12\epsilon^{2}J^{2}\Bigg(\frac{3\tA_{-}^{3}(n)\tA_{+}^{2}(n)}{2(K^{-1}-K)^2}
 \bigg(1+a(K)\bigg)\non\\
&&+\frac{\tA_{-}^{3}(n)\tA_{+}^{2}(n)}{(K^{-1}-K)D(K)}\Bigg).
\lb{hyp_rg-}
\eeqa
 The set of Eqs. \fr{hyp_rg+} and \fr{hyp_rg-} do not have the symplectic
 symmetry but is regularized by the folowing exponentiation procedure.
\beqa
\tA_{+}(n+1)&=&\tA_{+}(n)\exp\Bigg\{
\epsilon \frac{6J\tA_{+}(n)\tA_{-}(n)}{K-K^{-1}}\non\\
&&+\epsilon^2 12J^2\bigg(
    -a(K)\frac{3\tA_{+}^2(n)\tA_{-}^2(n)}{2(K-K^{-1})^2}
    +\frac{\tA_{+}(n)\tA_{-}(n)}{(K-K^{-})D(K)}
    \Bigg)\bigg\},         \lb{hyp_erg+}\\
\tA_{-}(n+1)&=&\tA_{-}(n)\exp\bigg\{
\epsilon \frac{6J\tA_{-}(n)\tA_{+}(n)}{K^{-1}-K}\non\\
&&+\epsilon^2 12J^2\bigg(
    +a(K)\frac{3\tA_{-}^2(n)\tA_{+}^2(n)}{2(K^{-1}-K)^2}
    +\frac{\tA_{-}(n)\tA_{+}(n)}{(K^{-1}-K)D(K)}
                \Bigg)\bigg\}.         \lb{hyp_erg-}
\eeqa
This regularized RG map has a constant of motion
$$
P\equiv\tA_{+}(n)\tA_{-}(n)=\tA_{+}(0)\tA_{-}(0),
$$
and is also exactly solvable. In terms of $P$, a
general solution of Eqs. \fr{hyp_erg+} and \fr{hyp_erg-}
is given as
\beqa
\tA_{+}(n)&=&\tA_{+}(0)\exp\bigg(Q(P;\epsilon)n\bigg),\non\\
\tA_{-}(n)&=&\tA_{-}(0)\exp\bigg(-Q(P;\epsilon)n\bigg),\non
\eeqa
where $Q(P;\epsilon)$ is a polynomial of $P$:
\beqa
Q(P;\epsilon)&\equiv&
\epsilon \frac{6JP}{K-K^{-1}}\non\\
&&+\epsilon^2 12J^2\bigg(
    \frac{-3a(K)P^2}{2(K-K^{-1})^2}
    +\frac{P}{(K-K^{-1})D(K)}
                \bigg).\non
\eeqa

\section{Four-dimensional Symplectic Map}
\qquad
Let us consider a coupled map of two symplectic maps \fr{nonmap} \\
$\bR^4\ni (x_n,y_n,x_n',y_n')\mapsto (x_{n+1},y_{n+1},x'_{n+1},y'_{n+1}):$
\beqa
x_{n+1}&=&\frac{\partial F}{\partial y_{n+1}},\quad
y_{n}  =\frac{\partial F}{\partial x_n},
\lb{4y}\\
x_{n+1}^{'}&=&\frac{\partial F}{\partial y_{n+1}^{'}},\quad
y_{n}^{'}=\frac{\partial F}{\partial x_n^{'}},
\lb{4y'}
\eeqa
where  $F$ is a gerenating function ,
\beqa
F(x_n,y_{n+1},x_n^{'},y_{n+1}^{'})
&=&x_n y_{n+1}+x_n^{'} y_{n+1}^{'}
  +\frac{1}{2}(y_{n+1}^2+y_{n+1}^{'2})\non\\
&&+\frac{1}{2}(Jx_n^2+J^{'}x_n^{'2})
  -2\epsilon\bigg(
    J\frac{x_n^4}{4}+J^{'}\frac{x_n^{'4}}{4}+ax_n^2x_n^{'2}
    \bigg).\non
\eeqa
The origin $(0,0,0,0)$ is an elliptic fixed point of \fr{4y} and \fr{4y'}
for $0< J<2$ and  $0< J^{'}<2$. Eliminating the $y$ and $y^{'}$
from Eqs. \fr{4y} and \fr{4y'}, we obtain coupled second order
difference equations:
\beqa
x_{n+1}-2\cos(\theta)x_{n}+x_{n-1}&=&
2\epsilon (Jx_n^3+2a x_n x^{'2}_n),\lb{second1}\\
x_{n+1}^{'}-2\cos(\theta')x_{n}^{'}+x_{n-1}^{'}&=&
2\epsilon (J^{'}x_n^{'3}+2a x_n^{2} x_n^{'}),\lb{second2}
\eeqa
where $\cos\theta=1- J/2$ , $\cos\theta'=1-J^{'}/2$.
Now, we concentrate on the case that the ratio of
frequencies of the leading order solution of \fr{second1} and \fr{second2}
,i.e.
$\theta/\theta'$ is an irrational number.
Then, straightforward calculations yield  perturbation solutions of
\fr{second1} and \fr{second2} as
\beqa
x_n^{(0)}&=&A\exp(i\theta n)+\cc,\non\\
x_n^{'(0)}&=&B\exp(i\theta^{'} n)+\cc,\non\\
x_n^{(1)}&=&\frac{-i}{\sin\theta}(3J|A|^2A+4a|B|^2A)
            n\exp(in\theta)+\cc+\nr ,\lb{sec1}\\
x_n^{'(1)}&=&\frac{-i}{\sin\theta^{'}}(3J^{'}|B|^2B+4a|A|^2B)
            n\exp(in\theta^{'})+\cc+\nr,\lb{sec2}
\eeqa
where $A\in\bC$ and $B\in\bC$ are integral constants.
In order to remove secular terms in \fr{sec1} and \fr{sec2}, we
introduce a RG transformation $A\mapsto \tA(n),B\mapsto \tB(n)$:
\beqa
\tA(n)&\equiv&A+\epsilon\frac{-i}{\sin{\theta}}(3J|A|^2+4a|B|^2)An,\non\\
\tB(n)&\equiv&B+\epsilon\frac{-i}{\sin{\theta^{'}}}
                  (3J^{'}|B|^2+4a|A|^2)Bn,\non
\eeqa
from which  a naive RG map is obtained as
\beqa
\tA(n+1)&=&\tA(n)\bigg\{1+\epsilon\frac{-i}{\sin{\theta}}
                          (3J|\tA|^2+4a|\tB|^2)\bigg\},\non\\
\tB(n+1)&=&\tB(n)\bigg\{1+\epsilon\frac{-i}{\sin{\theta^{'}}}
                     (3J^{'}|\tB|^2+4a|\tA|^2)\bigg\}.\non
\eeqa
To regularize this map, we scale the renormalized variables as
$$
\alpha(n)\equiv\frac{\tA(n)}{\sqrt{\sin\theta^{'}}},\qquad
\beta(n)\equiv\frac{\tB(n)}{\sqrt{\sin\theta}},
$$
where $ \sin\theta > 0$ and $ \sin\theta^{'} > 0$.
 Finally  we ``exponentiate'' the resultant naive RG map as
\beqa
\alpha(n+1)&=&\alpha (n)\exp\bigg\{
 \epsilon(-i)\bigg(3J\frac{\sin\theta^{'}}{\sin\theta}|\alpha(n)|^2
                 +4a|\beta(n)|^2\bigg)\bigg\}, \lb{rg4a}\\
\beta(n+1)&=&\beta (n)\exp\bigg\{
  \epsilon(-i)\bigg(3J^{'}\frac{\sin\theta}{\sin\theta^{'}}|\beta(n)|^2
                 +4a|\alpha (n)|^2\bigg)\bigg\},\lb{rg4b}\lb{rg4b}
\eeqa
It is easy to see Eqs.\fr{rg4a} and \fr{rg4b} have the sympletic symmetry
$$
d\alpha(n+1)\wedge d\alpha^{*}(n+1)+d\beta(n+1)\wedge d\beta^{*}(n+1)=
d\alpha(n)\wedge d\alpha^{*}(n)+d\beta(n)\wedge d\beta^{*}(n),
$$
and two constants of motion $|\alpha|^2$ and $|\beta|^2$.
\section{conclusion}
\qquad
We present
a regularized RG procedure to preserve the symplectic structure in
RG maps near elliptic and hyperbolic fixed points
of some symplectic discrete systems.
The regularization is accomplished by simple exponentiation of a
naive RG map and gives a symplectic RG map, which succesfully describes
a long-time asymptotic behaviour of the original system.
Without such regularization, a naive (truncated) RG map  fails
to describe a long-time behaviour especially near elliptic fixed points.
Furthermore, all regularized RG maps obtained here
have  constants of motion and can be exactly solvable, while
non-regularized  (nonlinear) RG maps have no conserved quantities
and are not solvable analytically.
As an application of the regularized RG map, we give an approximate
but analytical exprssion of the rotation number around an elliptic fixed
point. \\
It is easy to see that  the present regularization process is also
applicable to general  two-dimensional symplectic maps and higher
dimensonal symplectic maps, which have an elliptic fixed point with
incommensurate frequencies. However, construction of regularized RG
maps of general higher dimensional symplectic maps is still an open
 problem to be studied in future.

\section{Acknowledgements}
\qquad
The authors would like to thank Prof. T. Konishi and Dr. Y. Hirata
 for fruitful discussions.\\



\begin{thebibliography}{99}
\bibitem{nt-flow} For the corresponding Hamiltonian flows,
the trancated RG equation exactly preserves the symplectic structure
 with canonical variables $(\tA,\tA^{*})$ order by order and
 $|\tA|^2$ is an exact constant of motion.
\end{thebibliography}
\end{document}